# Axisymmetric Grazing-Incidence Focusing Optics for Small-Angle Neutron Scattering

D. Liu[1], M. V. Gubarev[2], G. Resta[3], B. D. Ramsey[2], D. E. Moncton[1,3], and B. Khaykovich[1*]

[1]Nuclear Reactor Laboratory, Massachusetts Institute of Technology, 138 Albany St., Cambridge, MA 02139, USA

[2]Marshall Space Flight Center, NASA, VP62, Huntsville, AL 35812, USA

[3]Department of Physics, Massachusetts Institute of Technology, 77 Massachusetts Ave., Cambridge, MA 02139, USA

**Abstract**

We propose and design novel axisymmetric focusing mirrors, known as Wolter optics, for small-angle neutron scattering instruments. Ray-tracing simulations show that using the mirrors can result in more than an order-of-magnitude increase in the neutron flux reaching detectors, while decreasing the minimum wave vector transfer. Such mirrors are made of Ni using a mature technology. They can be coated with neutron supermirror multilayers, and multiple mirrors can be nested to improve their flux-collection ability. Thus, these mirrors offer simple and flexible means of significantly improving existing and future SANS instruments. In addition, short SANS instruments might become possible, especially at compact neutron sources, when high-resolution detectors are combined with Wolter optics.

---

[*]Corresponding author. E-mail address: bkh@mit.edu

# 1. Introduction

Small-angle neutron scattering (SANS) is a widely used modern technique to study the micro-structure of various substances, with spatial sensitivity from nanometer to micrometer. SANS is especially powerful when applied to soft condensed matter, such as polymers, colloids, and biological macromolecules. SANS instruments are present at almost every neutron research center worldwide, from state-of-the-art spallation neutron sources to compact university-based facilities. SANS diffractometers serve a large community of users, including physicists, biologists, and chemical engineers. Therefore, improvements in their performance would have major impact. We show that significant improvements in both the signal rate and resolution of SANS instruments are possible by applying an innovative neutron focusing technique.

In SANS measurements, an incident beam of neutrons of certain wave vector $\boldsymbol{k}_i \pm \Delta \boldsymbol{k}_i$ is elastically scattered by a sample and scattered neutrons, of wave vector $\boldsymbol{k}_f \pm \Delta \boldsymbol{k}_f$, are counted. Here $|\boldsymbol{k}_i| = |\boldsymbol{k}_f| = k = 2\pi/\lambda$, $\lambda$ is the neutron wavelength and $\Delta \boldsymbol{k}$ represents uncertainty in the value and direction of the wave vector. Thus, the scattering cross-section is measured as a function of the wave vector transfer $\boldsymbol{Q} = \boldsymbol{k}_f - \boldsymbol{k}_i$. Here $Q = \frac{4\pi}{\lambda}\sin\frac{\theta}{2}$, and $\theta$ is the scattering angle between $\boldsymbol{k}_i$ and $\boldsymbol{k}_f$. The measured cross-section contains structural information about the sample. The typical range of $Q$ is $10^{-1}$ to $10^{-3}$ Å$^{-1}$, while scattering angles extend as low as milliradians.

Most existing SANS instruments use a collimation system that consists of two small apertures (often referred to as sample and source pinholes or slits), which together limit the size and the divergence of the beam. The beam is well collimated to minimize the direct-beam footprint on the detector, which is located several meters behind the sample. The smaller the footprint, the smaller the wave vector transfers $Q$ that can be achieved. Many SANS experiments require reaching the smallest possible $Q$. However, the two small apertures severely constrain the neutron flux on the sample and, hence, the signal rates on the detector. Flux is thus sacrificed to achieve minimal $Q$ ($Q_{min}$). The way to increase the flux without limiting the resolution was demonstrated long ago [1], by replacing the traditional sample aperture with focusing optics. The use of focusing optics increases the flux on a sample by collecting larger beam divergence. Also when the beam is focused on the detector, the footprint can be smaller than that in the traditional (pinhole) geometry, thus improving $Q_{min}$ [2]. In spite of the benefits, focusing optics is not commonly used at most SANS facilities, for reasons described below.

There are two kinds of neutron focusing optics based on either refraction or reflection. Refraction lenses are sometimes utilized at reactor SANS diffractometers, such as at NIST (USA), J-PARC (Japan) and HZB (Germany) [3-5]. Major drawbacks of such lenses are strong chromatic aberrations; the focal distance of a biconcave neutron lens changes as the second power of the neutron wavelength. Consequently, these lenses are not suitable for time-of-flight (TOF) SANS instruments. In fact, chromatic aberrations reduce the resolution even on continuous-beam SANS instruments since the beam is not perfectly monochromatic. A recently developed magnetic lens installed at ILL (France) offers a possible solution [6,7]. Magnetic lenses are also chromatic, but the new device reduces chromatic aberrations by modulating the magnetic field for different velocity neutrons. However, such lenses work only with polarized neutrons, reducing the available neutron flux. In addition, complicated magnets require constant support and maintenance during operations [8,9]. These disadvantages limit the utilization of magnetic lenses.

Reflection-based focusing devices, however, are free of chromatic aberrations and require little maintenance. A mirror-based SANS instrument is operational at JCNS (Germany) [1]. This instrument is equipped with a single toroidal mirror, 1.2 m long, coated with Cu [10]. Unfortunately, the instrument requires large samples or long counting times, presumably due to the limited collection efficiency of the mirror. Limitations of mirror performance arise from the exceedingly small critical incidence angle, above which almost no neutrons are reflected. However, the combination of optical designs and technology inspired by x-ray telescopes makes it possible to dramatically improve the performance of mirror-based SANS instruments, as demonstrated below.

We propose and analyze the use of axisymmetric Wolter-type mirrors, which are commonly utilized in x-ray astronomy and microscopy [11-14]. Neutron Wolter mirrors have been recently demonstrated by our group [15,16]. These mirrors can be made essentially free of optical aberrations, which could limit the instrumental resolution. In addition, several coaxial confocal mirrors can be combined together (nested) to increase the collection efficiency while keeping the length to a minimum. The mirrors can be made of Ni or even coated with neutron supermirror multilayers [17,18]. Our ray-tracing simulations demonstrate large potential improvements in the signal at the detector, as shown on Figure 1. The signal is predicted to increase by a factor of fifty or more, and $Q_{min}$ to decrease by a factor of two, for a currently operational SANS instrument equipped with a 0.4 m-long Ni ellipsoidal mirror.[1] The improvements stem from the ability of the optics to collect neutrons from a much larger solid angle than is possible with pinhole designs, when apertures define the beam size and divergence. Our optics can be easily optimized for other SANS instruments at both pulsed and reactor sources, leading to major enhancements in their performance.

In the following, we start by introducing principles and benefits of focusing, and then demonstrate design objectives and choices for axisymmetric focusing mirrors in a general case. Next, we will use a particular SANS instrument as an example to demonstrate the optimization of the optics, and improvements which stem from using it.

**2. Wolter optics for SANS**

The schematic of a traditional SANS instrument that uses two collimating apertures is shown in Figure 2(a). The instrument equipped with Wolter optics is shown in Figure 2(b). Although Wolter optics actually involve the use of pairs of confocal conical-section mirrors, only one ellipsoid mirror is shown for simplicity. When Wolter optics is used, the source is located at the first focus of the mirror. Incident neutrons are reflected by the mirror to the second focus, where the detector is located. The sample is positioned between optics and the second focus, as it is when the refractive optics mentioned earlier is used [2-6].

The benefits of using focusing optics for SANS are illustrated on Figure 3. Figure 3(a) shows the radius $R$ of the direct beam at the detector as a function of the optics-to-detector distance (ODD). The calculations are made with and without optics in the geometry of Figure 2. In changing the ODD, the optics is also changing in order to keep the focus at the detector. The smallest wave vector transfer shown on Figure 3(b) is $Q_{min} = \frac{4\pi}{\lambda}\sin(\frac{1}{2}\frac{R}{SDD})$, where $\lambda$ is the neutron wavelength and SDD is the sample-to-detector distance. The ODD is approximately equal to the SDD, since the distance between the optics and the sample is usually much smaller than that between the sample and the detector. With focusing optics, the direct beam size and hence $Q_{min}$ is significantly reduced.

---

[1] In this example, we used parameters of the new Extended-Q Small Angle Neutron Scattering (EQ-SANS) diffractometer which utilizes a broadband neutron beam at Spallation Neutron Source. SANS spectra have been calculated using a standard test sample, as described below.

**A. Optical design choices**

The collection of a diverging beam from a small source is achieved in geometrical optics by using a combination of conic sections, such as an ellipsoid, paraboloid or hyperboloid. We have considered three candidate geometries suitable for SANS: ellipsoid, paraboloid-paraboloid (PP) and hyperboloid-ellipsoid (HE) mirrors. Such optics can be routinely produced by our collaboration, using the technology originally developed for x-ray astronomy [14]. Schematic drawings of the optics are shown on Figure 4. The source and the detector are at the two focal points (O and $F_2$). For PP, the source is at the focal point of the upstream paraboloid, while the detector is at the focus of the downstream one. For HE, the source is at the focal point of the hyperboloid and the detector is at the focus of the ellipsoid. The second foci of the hyperboloid and ellipsoid coincide, as shown on Figure 4(c). For the ellipsoid neutrons are reflected once by the mirror before reaching the detector. In the two-reflection geometries, PP and HE, neutrons reflected by the first mirror reach the focus after being reflected by the second mirror.

The optics has been analyzed by ray-tracing, with the help of the software package McStas [19], which is routinely used for simulating neutron-scattering instruments and conducting virtual experiments [20]. In the quasi-classical approximation, which works well for slow neutrons, trajectories are represented as rays of geometrical optics. Interactions between the rays and mirrors are described mathematically using reflectivity curves, either measured or modeled. In general, neutrons are reflected from mirrors below a certain critical angle, above which the reflectivity drops to zero very quickly. Focusing mirrors must be optimized for a particular SANS instrument, since the performance of the optics depends on the mirror sizes, focal distances, and neutron energies, which determine critical angles. Therefore, the ray-tracing analysis was performed for an existing SANS diffractometer, EQ-SANS at the Spallation Neutron Source (SNS) [21,22].

Results of ray-tracing simulations of the three mirror geometries are shown in Figure 5. The source-to-optics distance is 4 m; the optics-to-detector distance is 9 m. Thus, the optical magnification is 2.25. The divergence of the neutron beam from the source aperture is determined by the neutron guide upstream of the source, $\theta \cong 1.73 m\lambda$ (mrad), where $m$ is the reflection parameter of the neutron guide ($m = 1$ for Ni, $m = 3.5$ for the considered beamline) and $\lambda$ is the neutron wavelength (Å). A point monochromatic source is used for simplicity, with the incident wavelength $\lambda = 4$ Å, and the optics is coated with Ni. In these simulations there was no sample between the source and the detector. The optics was considered to be ideal, without taking into account possible imperfections due to manufacturing, but including actual reflectivities.

The flux illuminating a sample is proportional to the solid angle subtended by the optics. The increase in the collected flux with the radius follows the corresponding increase of the solid angles shown on Figure 4(a). For the mirrors which consist of two or more optical components, only the solid angle illuminating the first component is captured, as neutrons must reflect from each component consequently between the source and focus. The sharp drops in the intensity seen on Figure 5 at about r = 0.04, 0.055 and 0.065 m appear when the incident angle becomes larger than the critical angle. Although PP mirrors can potentially capture larger flux, they are longer than the ellipsoid and, hence, more difficult to fabricate. Therefore, we analyze the performance of the elliptical mirror in more detail.

The performance of ellipsoidal mirrors is shown on Figure 6. The vertical axis represents the intensity gain, the ratio of the total neutron intensity on the detector with and without optics ($I_{optics}$ and $I_{no\_optics}$ correspondingly). With optics, the source-aperture radius is 5 mm. First, consider the traditional set-up (without optics), where both source and sample apertures have the same radii of 5 mm. The intensity gain in this configuration is represented by open symbols on Figure 6. From Figure 3(a), the direct beam without the optics is much larger than that with optics, meaning that $Q_{min}$ is improved. The improvement is by a factor of about 2 (see Figure 3(b)). To compare the intensity gain when $Q_{min}$ is the same, we used the following traditional configuration: the radius of the source is 2.5 mm, and of the sample aperture is 2 mm. Then, the direct-beam footprint at the detector in the traditional configuration is as small as it is when the optics is used (5 mm source radius). The solid symbols represent the result in this case, the increase in the intensity by almost two orders of magnitude. Furthermore, coaxial confocal mirrors of different radii can be nested to further increase the neutron intensity by increasing the solid angle of illumination by the beam. The corresponding intensity ratios are also shown on Figure 6. In this case, the horizontal axis represents the inner radius of the largest mirror. Significant increases in the signal are evident and grow with the mirror's size. For relatively small radius optics the neutron incident angles are so small that there is no wavelength dependence of the performance of the mirrors. Indeed, for the mirrors on Figure 6, the short wavelength cut-off is less than 2 Å, which is shorter than normal SANS wavelengths.

Mirrors larger than those shown in Figure 6 were also considered. The calculated performance of these mirrors is shown in Figure 7 at different neutron wavelengths. Ellipsoid and paraboloid-paraboloid mirrors were considered. Both Ni optics and multilayer optics have been simulated. Three types of mirrors are shown in each figure: one single Ni mirror, a pair of nested Ni mirrors, and a mirror coated with an $m = 2$ multilayer. Two different wavelengths, 8 Å and 13 Å, were used for the single-mirror case. In the traditional (no-optics) configuration both the source and sample aperture radii are 5 mm. Similar to Figure 5, the neutron flux increases with the radius of the optics. At a certain radius, the flux drops because the incident angle reaches the critical angle of the mirror coating. The decrease is not as sharp as that on Figure 5, since a round neutron source with $R_s = 5$ mm was used to simulate the actual configuration, rather than a point source.

**B. Optimization of mirrors**

Building on these results, we designed an elliptical Ni mirror for an instrument similar to EQ-SANS, and tested it by ray-tracing. Figure 1 shows calculated spectra for the instrument equipped with one such mirror. The mirror has the radius of $r = 0.09$ m, corresponding to the short-wavelength cut-off at 9 Å. The length of the mirror is 0.4 m. The focal distances are $L_1 = 4$ m and $L_2 = 9$ m. The neutron wavelength $\lambda$ is set to 13 Å. These parameters were chosen because the source-spectrum bandwidth at EQ-SANS is normally 9 to 13 Å, when EQ-SANS is configured to reach small wave vector transfers. The SDD ($L_2$) then is 9 m, as in our simulations. The beam-stop at the entrance aperture of the mirrors has the radius of 0.0825 m (see Figure 4(a)). We used a McStas component, which models the sample with the radius of gyration $R_g = 100$ Å and the transmission $T = 0.9$. In the configuration without optics, both source and sample aperture radii are 5 mm. Figure 1 shows that just a single relatively short Ni mirror results in a 50-fold gain in the signal intensity. (If the source size in the no-optics configuration were kept small enough to reach the same $Q_{min}$ as with the optics, the improvement of the intensity would become more than two orders of magnitude.) Increasing the length or adding one or two nested mirrors can further significantly increase the gain.

Particularly for EQ-SANS, the beam size and divergence are limited by several guard slits of $31 \times 31$ mm$^2$ between the source and the sample position. The guard slits limit the divergence of the beam, the size of the mirrors, and therefore the achievable flux gains. Open symbols on Figure 6 show that the intensity gain is small when the radii of the mirrors are limited to less than 16 mm by the guard slits, relative to the standard configuration when both source and sample apertures are of 5 mm radius. If guard slits were removed, the full beam divergence from the $m = 3.5$ guides upstream of the source slit could be collected by the optics, resulting in the intensity increase as shown on Figure 1.

## 3. Discussion

Our results suggest that axisymmetric focusing mirrors can lead to dramatic improvements in the performance of SANS instruments, both at accelerator- and reactor-based facilities. We discuss below the design and use of these optics in more detail.

### A. Collection of a divergent beam

The flux illuminating a sample is proportional to the solid angle subtended by the optics. This solid angle, which increases with the size of the mirrors, is much larger than that subtended by a standard sample aperture. Thus, impressive gains are achieved by using the mirrors. The gains in the signal shown on Figure 1 result from the increase of the size of the sample illuminated by the beam. If the acceptable sample size were a limiting factor, it could be a parameter of the design. (If the sample can be moved away from the mirrors and closer to the focus, the size of the illumination area would become smaller, leading to the increased flux density and the signal-to-noise ratio. However, the increase in the signal would come at the expense of the resolution and $Q_{min}$.) The size of the mirrors is limited by the critical angle, which leads to the sharp cut-off radius above which the mirrors become ineffective. The neutron-flux gain does not depend on the wavelength below the cut-off radius, but the cut-off radius increases with the wavelength. Therefore, additional improvements of the performance of the mirrors can come from coating reflecting surfaces with neutron supermirror multilayers. Such coatings increase the critical angle, allowing for mirrors with larger cut-off radii. In general, for long-wavelength neutrons ($\lambda \geq 8$ Å) Ni mirrors are sufficient, while multilayer coating might be needed for shorter wavelengths. In order to take full advantage of the axisymmetric mirrors, the beam from the source should be divergent enough to illuminate large mirrors, ideally close to the cut-off size for the shortest wavelength. The divergence of the beam is determined by the optics upstream of the source aperture. Neutron guides that are often installed there should provide enough divergence to illuminate the mirrors. The position of the mirrors relative to the source and detector, and thus the magnification, are parameters of the design, and should be optimized for any given instrument.

Optical aberrations are weak for the short mirrors used here, but longer mirrors might introduce noticeable distortions of the image [23]. Indeed, the magnification of an elliptical mirror is different for every incident ray depending on its angle with the optical axis. Therefore, the image at the focus is distorted. Wolter has shown that to avoid such aberrations, even number of reflections must be used [11]. Therefore, for the SANS instruments, which allow long optics in front of a sample, two-reflection geometry might be beneficial, such as PP and HE described earlier. The surface roughness of the mirrors leads to diffuse scattering, which can contribute to the background. The mirrors are polished to less than 0.5 nm root-mean-square roughness amplitude, better than most commercial neutron guides, thus limiting the adverse diffuse scattering. The effect of the roughness will be studied in detail in the future. The experience from the mirror-based SANS instrument shows that the roughness effects can be tolerated [1].

**B. Development of future mirror-based SANS instruments**

In traditional SANS instruments, long vacuum tanks occupy the space between a sample and a movable detector. As we discuss below, focusing mirrors might eliminate the need for such a movable detectors and lead to simpler and shorter SANS instruments. Normally, users need to adjust SDD to change the Q coverage. A typical SANS measurement may use 2 to 3 different SDDs to cover the required Q range. To reach the smallest required $Q_{min}$, long SDD's are used, thus requiring long evacuated detector tanks. For instance, EQ-SANS has a 10 m detector tank [21], GP-SANS at HFIR (High Flux Isotope Reactor, ORNL) has a 20 m detector tank [24], and NG3 SANS in NIST has a 13 m detector tank [25]. These huge vacuum tanks are expensive, cumbersome, and require a lot of maintenance. By using optics, SANS instruments might be able to use shorter constant SDD's to reach the smallest $Q_{min}$. As shown on Figure 3(b), when mirrors replace the pinhole sample aperture, $Q_{min}$ becomes a constant independent of SDD. The length of the instrument is then determined by the focal length of the optics and the resolution is limited by the detector's pixel size. Modern multi-channel-plate detectors have pixel sizes of less than 40 μm, while offering excellent time resolution [26,27]. Hence, the optics offers at least two advantages. First, users could measure a large Q range in a single configuration, saving experimental time. Second, detector tanks could be made significantly shorter and without moving parts, therefore cheaper and easier to maintain.

The combination of novel optics and detectors will lead to revolutionary changes in SANS instruments, including extended Q-range and very fast (seconds) measurements of standard samples. New modalities of SANS could become more widely available, for example *in situ* studies of kinetic processes such as chemical reactions (such measurements are only rarely performed now because of low signal rates). Short mirror-based SANS instruments might be especially useful for compact accelerator-based sources, such as LENS [28] and others.

## 4. Conclusions

We have demonstrated multiple advantages of using novel axisymmetric focusing mirrors for SANS instruments. Using the parameters of the new EQ-SANS instrument at the SNS as an example, we predicted, that the optics could improve the signal by a factor of 50 or more, while decreasing the minimum wave vector transfer almost two-fold. Such optics, as routinely manufactured by our collaboration, can lead to dramatic improvements in the performance of SANS instruments at almost any neutron facility. We now have a ray-tracing tool to simulate mirrors for various facilities. The ray-tracing can also take into account imperfections due to manufacturing of mirrors. The analysis of such imperfections will be explored in a future publication.

## 5. Acknowledgements

We are grateful to Professor Jeffrey Gordon of Ben-Gurion University (Israel) and Dr. David Mildner of NIST for discussions. Research supported by the U.S. Department of Energy, Office of Basic Energy Sciences, Division of Materials Sciences and Engineering under Award # DE-FG02-09ER46556, # DE-FG02-09ER46557 (Wolter optics studies) and by National Science Foundation under Award # DMR-0526754 (construction of Neutron optics test station and diffractometer at MIT).

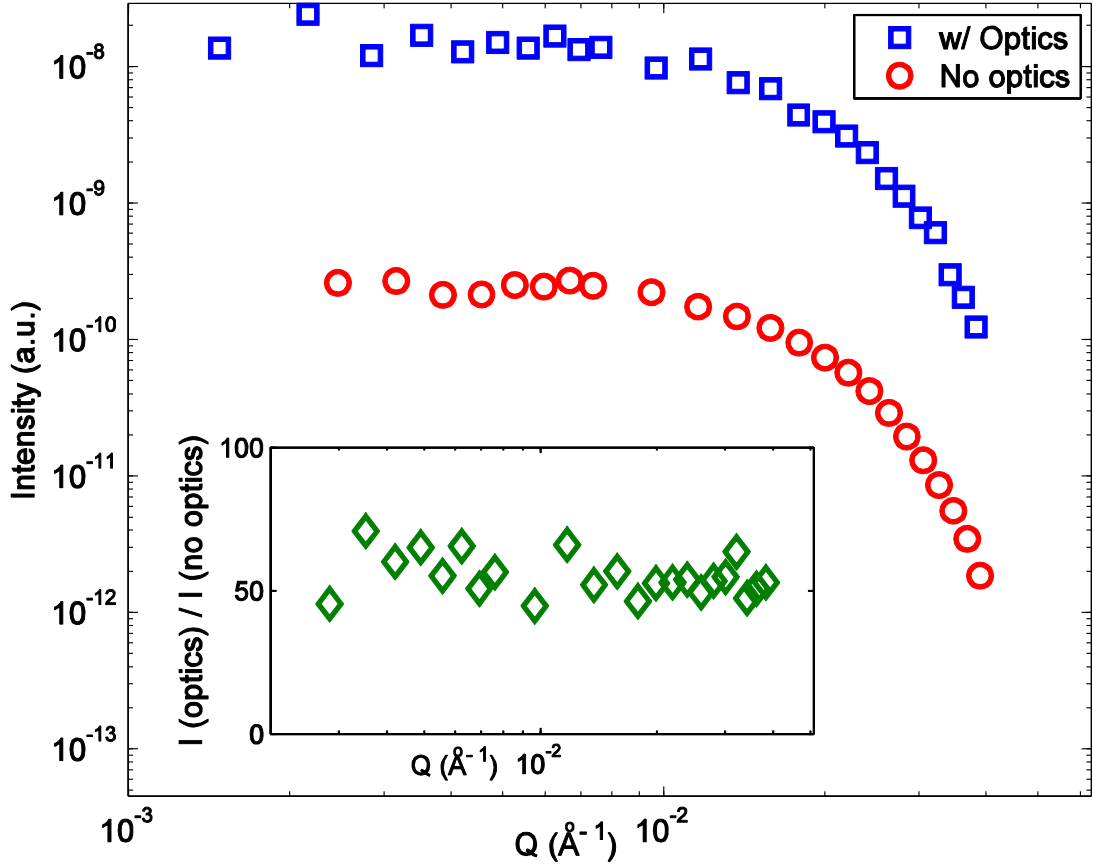

**Figure 1** (color online). Simulation results of SANS spectra with and without focusing mirrors. The test sample has the radius of gyration $R_g = 100$ Å, sample transmission $T = 0.9$. Blue squares are the SANS spectrum simulated with the designed optics and with a source aperture of 5 mm radius. Red circles are the SANS spectrum simulated for a standard (pinhole) geometry with both source and sample apertures of 5 mm radius. The incident neutron wavelength $\lambda = 13$ Å. **Inset**: The ratio of intensities of the two spectra.

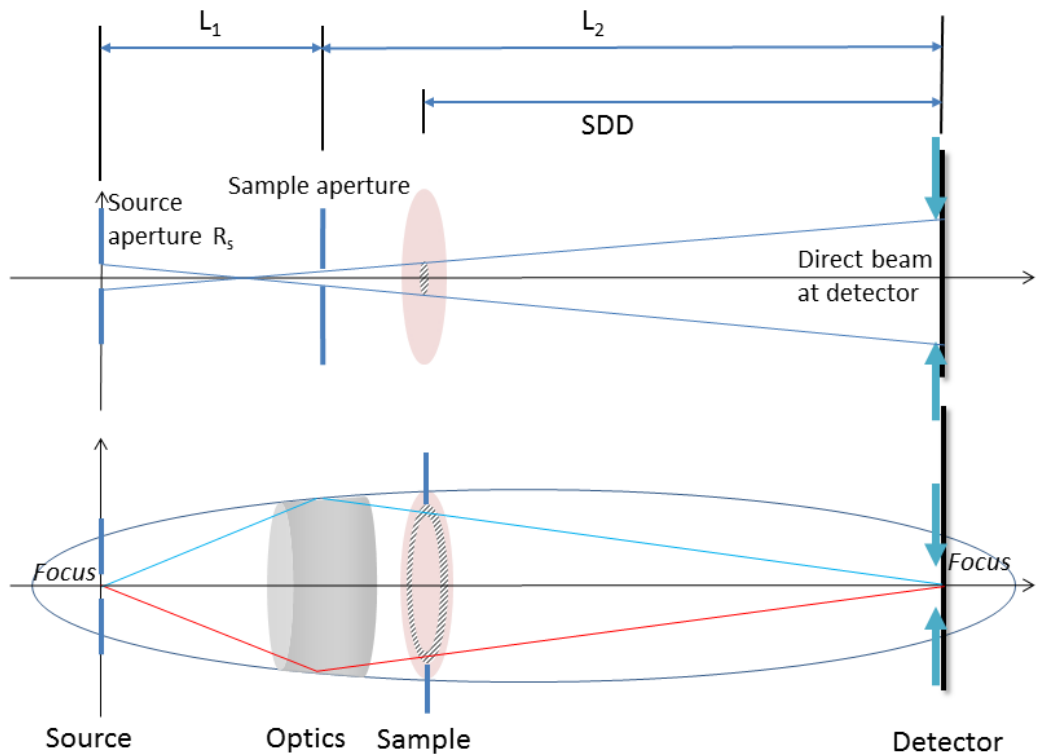

**Figure 2** (color online). Schematic layout of a traditional pinhole SANS (top) and a SANS with focusing optics (bottom). $R_s$ is the radius of source aperture. SDD denotes the sample-to-detector distance. $L_1$ is the source-to-optics distance (SOD). $L_2$ is the optics-to-detector distance (ODD). Magnification is defined by $M = L_2/L_1$. The two arrows indicate the direct beam spot at the detector. In the presence of focusing optics, the radius of the direct beam at the detector is $R = R_s M$. Only neutrons reflecting from the first mirror reach the detector because the center of the optics is blocked at the entrance aperture as shown schematically on Figure 4. Therefore, the beam illuminates only the shaded area of the sample. This area is larger with optics than without it.

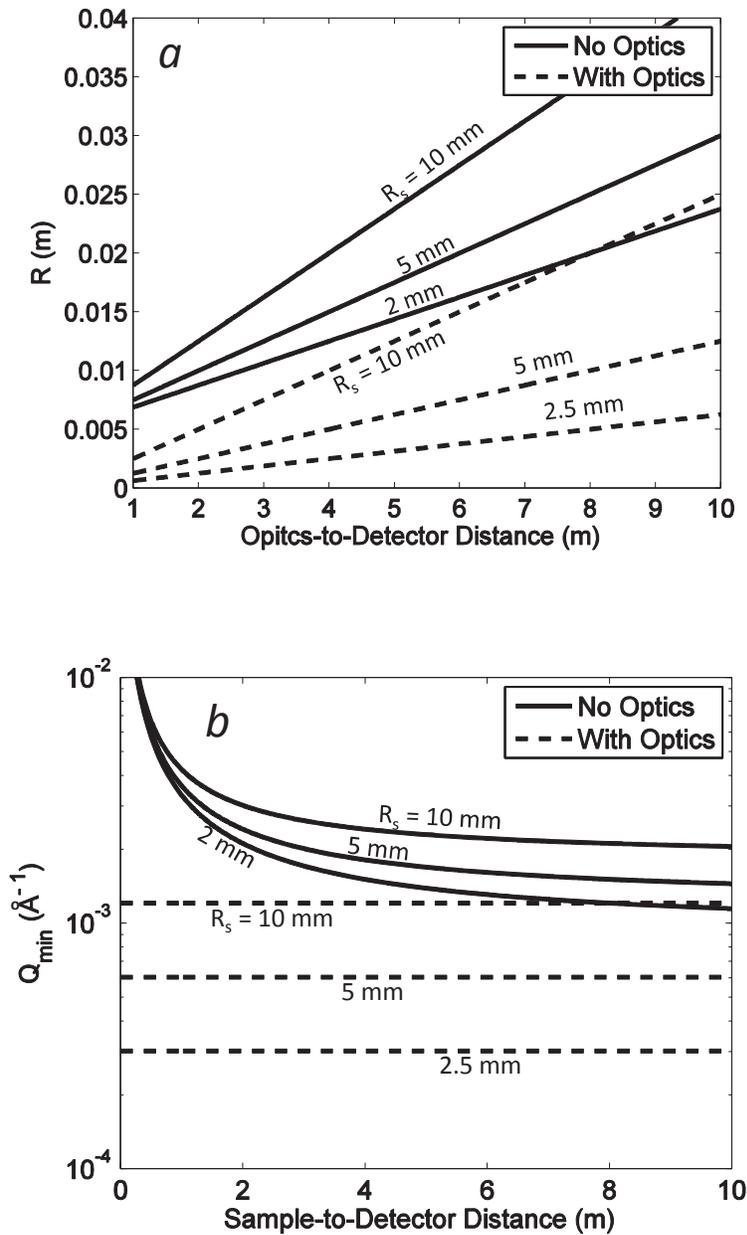

**Figure 3**. (a) Direct-beam radius at the detector (R) vs optics-to-detector distance and (b) the smallest wave vector transfer ($Q_{min}$) vs the sample-to-detector distance; see Figure 2 for the schematic. Dashed lines represent the setup with focusing optics. Solid lines represent the configuration without optics. The source-to-optics distance is constant, $L_1$ = 4 m. The sample aperture is 5 mm radius. The neutron source aperture, $R_s$ = 10, 5, or 2.5 mm radius. The incident neutron wavelength $\lambda$ = 13 Å. In practice, ODD and SDD are not smaller than about 1 m.

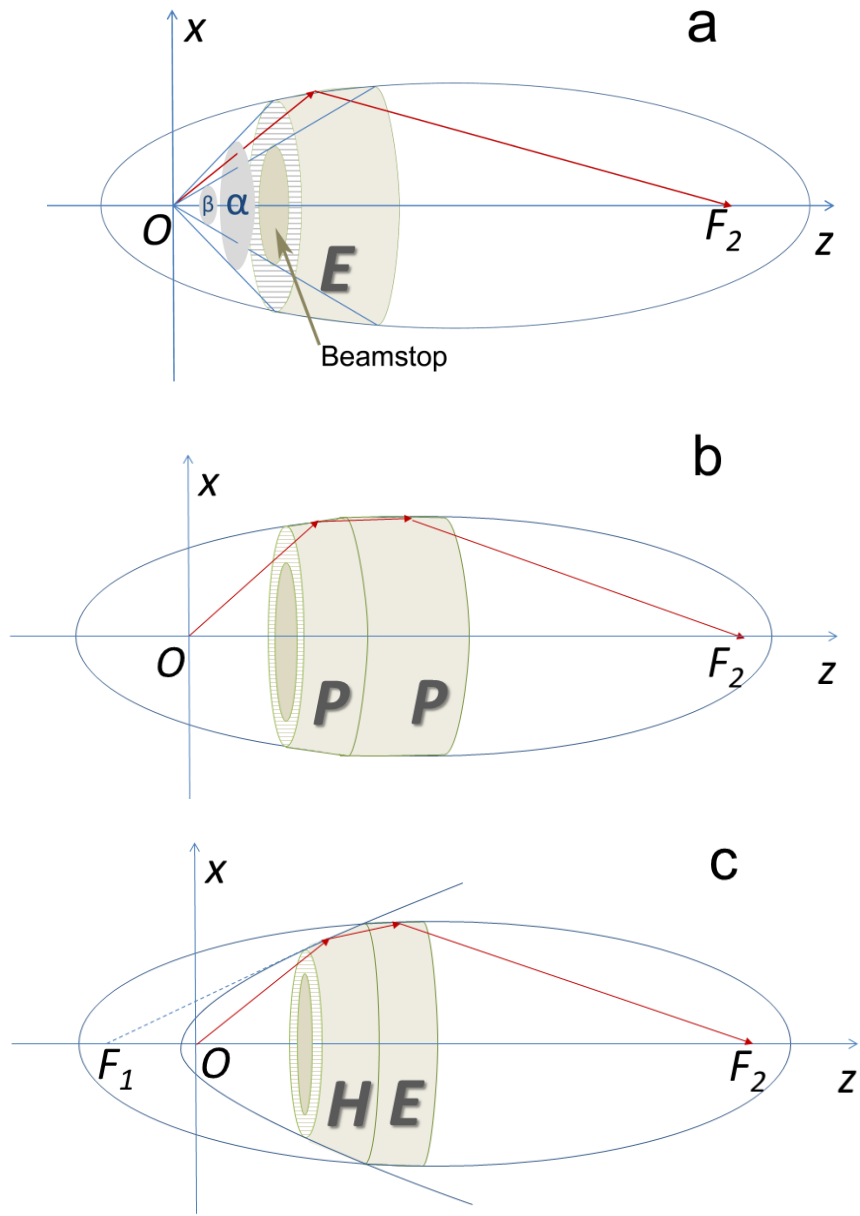

**Figure 4** (color online). Schematic drawings of three types of optics. "E", "P", and "H" denote ellipsoid, paraboloid and hyperboloid respectively. "O" denotes the origin, where the source is located and "$F_2$" is the focus with the detector. (a) Ellipsoid mirror. Neutrons reflect only once. The shaded area indicates the captured solid angle, which is the difference of solid angles α and β. (b) Paraboloid-Paraboloid (PP) mirror. O and $F_2$ are the foci of the two paraboloids. Neutrons reflect twice in this mirror. (c) Hyperboloid-Ellipsoid (HE) mirror. $F_1$ is the common focus of the hyperboloid and the ellipsoid. O is another focus of the hyperboloid. Neutrons reflect twice in this mirror. The circular beam-stop, shown on all drawings, blocks the center of the optics at the entrance aperture. The diameter of the beam-stop is such that it allows only neutrons, which intersect the first mirror, to continue. Additional shielding will stop neutrons with trajectories outside the mirrors.

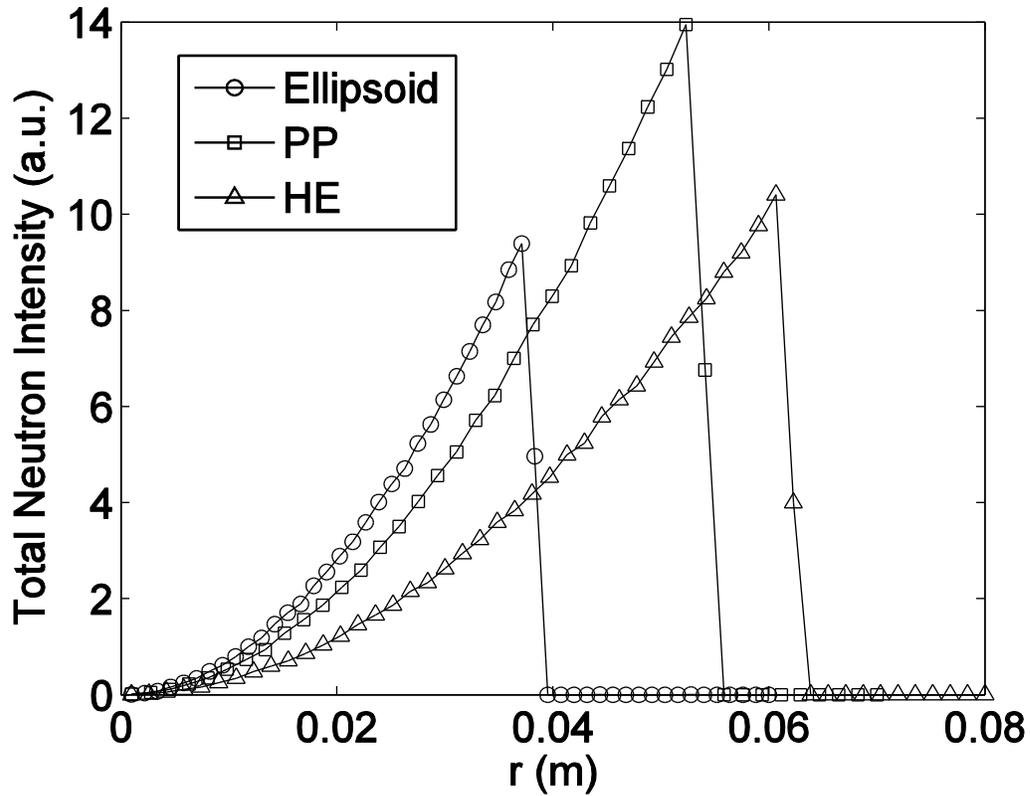

**Figure 5**. Comparison of the performance of three kinds of mirrors, when the magnification M = ODD/SOD = 2.25. X-axis represents the radii of the mirrors (at the middle). Y-axis represents the total neutron intensity at the detector. The circles, squares and triangles represent the ellipsoid, paraboloid-paraboloid (PP) and hyperboloid-ellipsoid (HE) mirrors respectively. A point neutron source is used for simplicity. The mirrors are coated with Ni (m = 1). The incident neutron wavelength $\lambda$ = 4 Å. The length of the ellipsoid mirror is 0.4 m. The total length of the PP optics is 1.3 m (the first paraboloid is 0.4 m, while the second one is 0.9 m = M×0.4 m). For HE, the total length of the optics is 0.8 m (both hyperboloid and ellipsoid mirrors have the same length of 0.4 m).

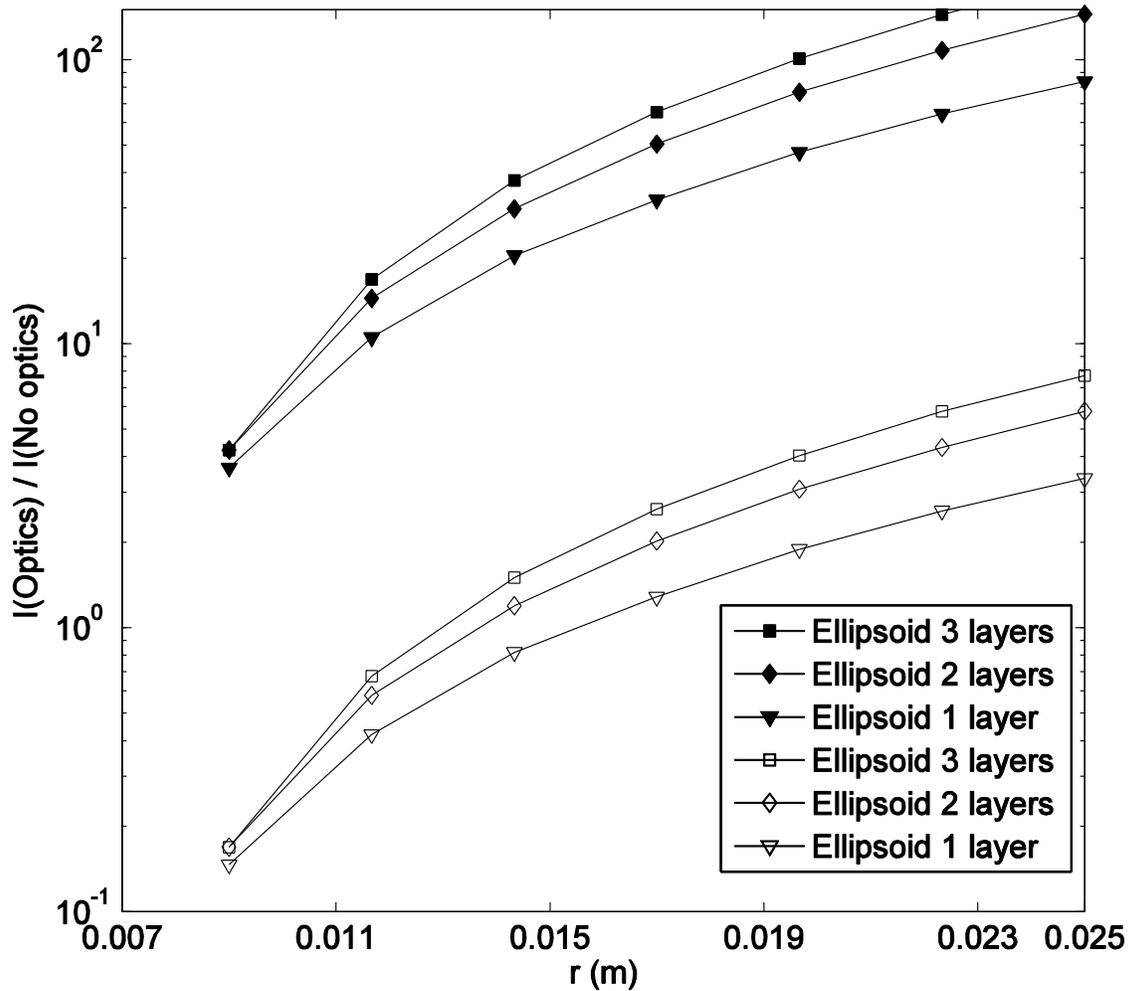

**Figure 6.** Neutron intensity ratio vs radius of the optics. The radius is taken at the middle of the ellipsoid. In the case of nested mirrors, the radius is that of the largest ellipsoid. The intensity ratio with mirrors and with pinholes is calculated at two different pinhole configurations. Solid symbols represent the following traditional (no-optics) configuration: the source radius $R_s$ = 2.5 mm and the sample aperture radius $R$ = 2 mm; open symbols represent the configuration with equal source and sample apertures: $R_s = R$ = 5 mm. In the configuration with optics, the source radius is always 5 mm.

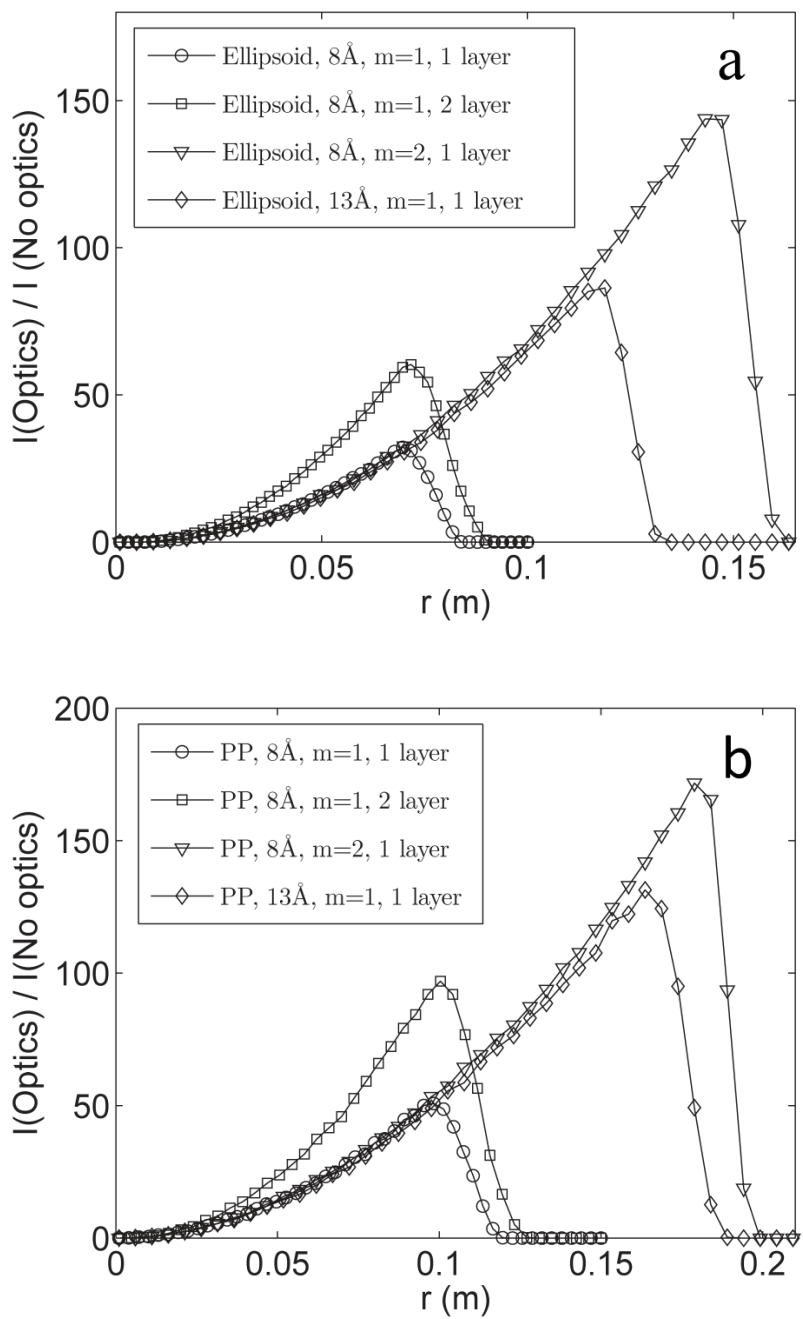

**Figure 7.** Neutron intensity ratio vs radius of the optics for large optics. (a) Ellipsoid mirror. (b) Paraboloid-parabolid (PP) mirror. The radius is taken at the middle of the mirror. The ratio is between the total neutron intensity with and without mirrors. In the pinhole configuration, the source and sample apertures are the same, of 5 mm radius. The ellipsoid mirror is 0.4 m long. The first paraboloid part of the PP mirror is 0.4 m long.